# TURBULENCE AND MIXING IN THE EARLY UNIVERSE

**Carl H. Gibson**

Departments of Mechanical and Aerospace Engineering and Scripps Institution of Oceanography, University of California at San Diego, La Jolla CA 92093-0411, USA, cgibson@ucsd.edu, http://www-acs.ucsd.edu/~ir118

**Abstract**--The role of turbulence and turbulent mixing in the formation and evolution of the early universe is examined. In a new quantum-gravitational model with fluid mechanics, it is suggested that the mechanism of the hot big bang is functionally equivalent to the mechanism of turbulence, where an inertial-vortex force at Planck scales matches the Planck gravitational force and drives the formation of space-time-energy and the formation of more Planck particles, more spinning Planck-Kerr particles, and a big bang turbulence cascade to larger scales before cooling to the strong force freeze out temperature. Temperature fluctuations between the Planck temperature $10^{32}$ K and strong force temperature $10^{28}$ K are mixed by turbulence with Reynolds numbers up to $10^6$ to give a Corrsin-Obukhov spectral form. Gluon viscous forces then damp the turbulence and help cause an exponential inflation of space that fossilizes the turbulent temperature fluctuations by stretching them beyond the horizon scale of causal connection $ct$, where $c$ is light speed and $t$ is time. These fossil temperature turbulence fluctuations seed anisotropies in the nucleosynthesis of light elements, causing density fluctuations that seed the first formation of gravitational structure in the matter dominated hydrogen-helium plasma as proto-voids at galaxy supercluster scales. Evidence of the proposed big bang turbulence event and first gravitational structure formation is provided by the spectral form of the cosmic microwave temperature fluctuations, which have been misinterpreted as sonic.

*Keywords: Turbulence, Turbulent Mixing, Cosmology, Astrophysics*

## INTRODUCTION

Understanding the formation and evolution of the early universe is complicated by a failure of general relativity theory and quantum mechanics at the initial stages of the big bang, during the epoch of quantum gravitational dynamics (QGD). Both GR and QM theories exhibit singularities as temperatures, densities and length scales approach the extreme Planck scale values determined by four parameters: light speed c, Planck's constant h, Newton's gravitational constant G, and Boltzmann's constant k. Some of the problems are resolved by M-superstring theories (MS), in which the fundamental particles of nature such as photons and electrons are taken to be multidimensional stringlike objects on Planck length scales that are distinguished by their different oscillation modes, and where the least massive black holes (Planck particles) undergo a phase change to zero mass photons, Greene 1999. However, fluid mechanics (FM) theory and the possibility of turbulence in the QGD epoch have not been considered, so this is the main goal of the present paper. The conclusion is that not only is the QGD epoch turbulent, it is QGD turbulence that powered the big bang.

Linear perturbation methods that may be very successful for many problems break down for all three standard physical theories (GR, QM, and MS) as the Planck scales of the big bang are approached, just as they do for FM under conditions where flows become turbulent. The QGD flow is highly nonlinear and fluid mechanics effects cannot be ignored. It is suggested here that a wide variety of outstanding problems in astrophysics and cosmology are resolved if fluid mechanics theory is properly included to account for viscous, diffusive, and turbulent transport processes, Gibson 1996, 2000.

Unfortunately, fluid mechanical considerations rarely enter discussions of the standard cosmology such as Peacock 2000, where the terms *viscosity, diffusivity* and *turbulence* are absent from the book's index and virtually absent from the text. Cosmic fluids are generally taken to be inviscid and their flows ideal, linear and acoustical. These assumptions are quite unjustified in the turbulent QGD epoch and in the very hot radiation-dominated epoch after the big bang and inflation before about $10^{10}$ s when viscosity forces due to the powerful radiation greatly exceed both the inertial-vortex forces and the gravitational forces. Gibson 1996 suggests that the first time after inflation when viscous, gravitational, and inertial-vortex forces of the expanding Hubble flow all became of comparable magnitude is about t = $10^{12}$ s = 30,000 years after the big bang. Weak turbulence possibly appears, but it is rapidly damped by buoyancy forces of the first gravitational structures: proto-voids between protosuperclusters that form as the cooling plasma fragments, nucleated by non-acoustic density minima remnant from the QGD epoch. The voids fill up with the strongly diffusive, "weakly interacting massive particle" (WIMP), nonbaryonic fluid that constitutes most of the matter of the universe according to most cosmological models. With further cooling, smaller proto-voids form and fill at proto-galaxy-cluster and proto-galaxy scales, down to $10^{20}$ m when the plasma turns to gas at t = $10^{13}$ s.

Fragmentation of the nonbaryonic dark matter occurred *after* the plasma to gas transition, not before as assumed in cold dark matter (CDM) models. The hypothetical dark matter is cold to reduce its Jeans condensation scale by reducing its sound speed. Gibson 1996 shows that CDM theories are unfortunate misconceptions arising from the incorrect Jeans 1902 gravitational instability theory. The Jeans scale acoustic criterion for the formation of structure by gravity is incorrect due to unjustified simplifications of the fluid mechanics problem, which leave out viscous and turbulence forces and the effects of diffusivity.

The nonbaryonic dark matter is very weakly collisional, so its particles have very long mean free paths between collisions and very large diffusivities. They cannot possibly condense during the plasma epoch because their gravitational diffusive scales are larger than the horizon. Observations of dense galaxy cluster haloes of scale $3 \times 10^{22}$ m indicate that the fragmentation time of the nonbaryonic dark matter occurred after the formation of galaxies, not before. The fragmentation scale was between the dense galaxy cluster scale of $10^{23}$ m and the outer galaxy halo scale of $10^{22}$ m, smaller than the supercluster scale $10^{24}$ m but larger than luminous disk galaxy scales of $10^{21}$ m. Galaxy cores at $10^{20}$ m represent the nonexpanded fossils of protogalaxies. Most of the original galaxy matter is dark (about 99.9%), and has diffused away from the cores to form $10^{22}$ m outer dark halos of the WIMP fluid (which behaves like neutrinos, and may be neutrinos) and $10^{21}$ m inner dark halos of baryonic dark matter in the form of dense proto-globular-star-cluster (PGC) clumps of frozen hydrogen-helium small-planet-mass objects.

The linear perturbation stability analysis of the inviscid Euler equations assumed by Jeans 1902 is rejected by Gibson 1996 in favor of new gravitational instability criteria determined by viscous and inertial-vortex forces balancing self gravitational forces for the primordial baryonic matter (hydrogen-helium plasma), or by diffusion velocities balancing gravitational velocities for the nonbaryonic matter (diffusive, dark and unknown). The Jeans criterion for gravitational instability states that density perturbations are stable at scales smaller than the Jeans length scale

$$L_J = V_S/(\rho G)^{1/2}$$

where $V_S$ is the speed of sound and $\rho$ is the density. Because the sound speed in the hot plasma epoch was of order the speed of light, the Jeans scale $L_J$ for the plasma was larger than the horizon scale, so by Jeans' criterion no structure could form except for some sort of hypothetical cold (meaning small $V_S$) dark matter. The new instability length scales introduced by Gibson 1996 are the viscous Schwarz scale

$$L_{SV} = (\nu \gamma / G)^{1/2}$$

the turbulent Schwarz scale

$$L_{ST} = \varepsilon^{1/2}/(\rho G)^{3/4}$$

and the diffusive Schwarz scale

$$L_{SD} = (D^2/\rho G)^{1/4}$$

where $\nu$ is the kinematic viscosity, $\gamma$ is the rate of strain, $\varepsilon$ is the viscous dissipation rate, and D is the particle diffusivity. Both $L_{SV}$ and $L_{ST}$ are estimated to be smaller than the horizon scale in the plasma epoch after about $10^{12}$ s by Gibson 1996, 2000, so gravitational fragmentation began despite the violation of Jeans' criterion.

The diffusive Schwarz scale $L_{SD}$ of the nonbaryonic dark matter exceeds the horizon length $L_H = ct$ in the plasma because its particles are weakly collisional, so that the mean free path for collisions $L_C$ is large and the velocity of the particles v is also large, giving a very large diffusivity D $\approx L_C v$. The CDM theory problem is compounded by another misconception appearing in many cosmological models that the diffusivity of fluids with weakly collisional particles is small rather than large. It is assumed that a sphere of "cold dark matter" particles will not expand by diffusion because the particles are gravitationally bound, but are so nearly collisionless that the "relaxation time" ($R^2/D$) is essentially infinite (of order R/v N/lnN $\approx 10^{88}$ s for N particles like neutrinos) corresponding to zero diffusivity D for finite R, Binney and Tremain 1987, p190. A counter example to this "collisionless fluid mechanics" model is a sphere with radius $R < L_{SD}$ of initially motionless, weakly collisional particles of uniform density $\rho$. All of the particles will fall to the center of the sphere after a free fall period $\tau_{FF} = (\rho G)^{-1/2}$ where the density will go to infinity and the particles will be forced to exchange momentum and randomize their velocities. Thus the size of the sphere and the effective diffusivity $D_E = R^2/\tau_{FF} = (RMG)^{1/2}$ will monotonically increase with time until $D_E$ matches the particle diffusivity D and R matches the Schwarz diffusive scale $L_{SD}$, where M is the total mass.

This idea that fluids composed of weakly collisional particles are nondiffusive is widely accepted, but highly questionable. How can D approach zero as $L_C = 1/n\sigma$ approaches infinity as the collision cross section $\sigma$ of the particles approaches zero for nonzero number density n and velocity v? Any concentration of gravitationally unbound particles will become less concentrated with time because the particles move independently in random directions. If the particles are gravitationally bound, as in the imaginary galaxy mass fragments of CDM required to get structure in Jeans-CDM models, the cloud will become more and more spherical and the particle paths will become more and more focused on the center with time constant $\tau_{FF}$. Collisions are forced by focusing to occur at the core of the sphere, giving a monotonic increase in the radius R until it reaches $L_{SD}$, where gravitational velocities match diffusional velocities as before. From the scale of dark matter halos in dense galaxy clusters it is possible to estimate the diffusivity of the material to be about $10^{28}$ m$^2$ s$^{-1}$, much larger than the diffusivity of any baryonic particle gas.

Standard cosmological models that rely on the Jeans theory predict a gravitational collapse of gas into cold dark matter (CDM) potential wells beginning at about 5 billion years after the big bang to form the first galaxies, which then collapse to form galaxy clusters and superclusters. The fluid mechanical criteria of Gibson 1996 give very different scenarios. According to these criteria, no fragmentation of any nonbaryonic matter occurs in the plasma epoch. The strong diffusivity of this nearly collisionless material gives condensation scales $L_{SD}$ larger than the horizon scale $L_H$. Condensation is impossible whether or not it is "cold" so $L_J < L_H$. Jeans' criterion also fails for the ordinary baryonic matter, so that galaxies and superclusters of galaxies were already fragmented before the plasma to gas transition at 300,000 years. None of these large structures have ever collapsed.

The first gravitational collapse (density increase) in the universe occurred when the gas appeared from the cooling plasma and the viscosity decreased by a factor of a trillion, to $10^{13}$ m$^2$ s$^{-1}$, giving planetary mass "primordial fog particles" (PFPs) that comprise the baryonic dark matter, Gibson 1996. The proto-galaxies simultaneously fragmented at the Jeans mass (for reasons different from those of the Jeans 1902 theory) to form $10^{36}$ kg proto-globular-star-clusters (PGCs), each containing about $10^{12}$ dark, planetary-mass, gas blobs. Most of these PGCs and their PFPs have remained dark and are now very cold and the PFPs frozen, except when neighboring galaxies merge and young globular star clusters pop out of the dark because tidal forces trigger an accretional cascade of the PFPs to form stars. The Hubble space telescope (HST) has the resolution to prove that these young globular clusters exist, and have ages of only a few million years, Larsen et al. 2001. Their densities of $10^{-17}$ kg m$^{-3}$ match the protogalaxy density, and the population of star types in the clusters matches that of the ancient globular star clusters observed in outer galaxy halos, with ages about 13-15 billion years matching that of the universe. No fluid mechanical scenario has been put forward, other than the Gibson 1996 model, that explains how hundreds of globular star clusters with the standard GC mass and density can suddenly appear out of the dark with a density that matches that of protogalaxies at age 0.0003 billion years. The HST resolution has revealed similar young globular star clusters in starburst and merging galaxies NGC 1705, NGC 1569, the Antennae, NGC 7252 and NGC 3256.

The best evidence that the baryonic dark matter of galaxies is in the form of planetary mass objects is provided by the quasar microlensing observations of Schild, 1996. Schild observed that the twinkling frequency of the gravitationally lensed images of a quasar corresponded to small planetary mass objects such as the earth, a million times less massive than a star. Thus the baryonic mass of at least one galaxy is dominated by point mass objects like PFPs.

Random density fluctuations are required to provide nuclei for gravitational structure formation at all scales, and it is generally assumed that these fluctuations were formed by some sort of chaos in the big bang event. Turbulence is not considered in standard textbooks about cosmology such as Weinberg 1972, Silk 1989, Kolb and Turner 1990, Peebles 1993, Padmanabhan 1993 and Rees 2000, despite the manifest nonlinearities of the initial conditions. In this paper a turbulent hot big bang model with inflation is proposed, and the predictions of the model are compared to observations of the cosmic microwave background spectrum.

We must first carefully define what we mean by the term turbulence.

### DEFINITION OF TURBULENCE

The proper use of fluid mechanics in studies of the early universe has been greatly complicated by a widespread disagreement about the definition of turbulence. What are the fundamental properties of turbulent flows, how is turbulence produced and driven, and what is the direction of the turbulence cascade? Turbulence is often either not defined by authors using the term, or it is defined in such a vague way that almost any random flow will qualify. In the present paper a narrow definition of turbulence is used, based on the conservation of momentum equation in the form

$$\frac{\partial \vec{v}}{\partial t} = -\nabla B + \vec{v} \times \vec{\omega} + \vec{F}_V + \vec{F}_G + \vec{F}_C + \ldots; \quad B = v^2/2 + p/\rho$$

where the usual nonlinear term of the Navier Stokes equations has been written as

$$(\vec{v} \cdot \nabla)\vec{v} = \nabla(v^2/2) - \vec{v} \times \vec{\omega}$$

so that the Bernoulli group $B = p/\rho + v^2/2$ can be isolated from the inertial-vortex force per unit mass

$$\vec{v} \times \vec{\omega}$$

that is considered to be a necessary part of any turbulence definition. The enthalpy $p/\rho$ and kinetic energy per unit mass $v^2/2$ adjust to each other locally to keep B nearly constant along streamlines in turbulent flows, eliminating its gradient in the momentum equation.

The inertial-vortex force is the term that always produces turbulence when it is larger than all of the forces that tend to damp it out such as viscous, gravitational, Coriolis, and electromagnetic forces given in the momentum equation. The inertial-vortex force causes vortex sheets to be unstable because it is perpendicular to the vortex sheet, so that any perturbation of the sheet will be amplified in the direction of the perturbation to form eddies, as shown in Figure 1a.

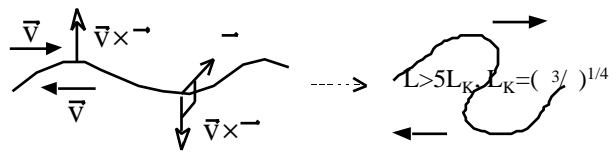

**Fig. 1a. Basic mechanism of turbulence formation**

The usual criterion for the formation of turbulence is whether the Reynolds number is above a critical value, but it is easy to show that the Reynolds number is the ratio

$$Re = \vec{v} \times \vec{v} / \vec{F_V}$$

formed by dividing the inertial-vortex force by the viscous force. Similarly, other critical ratios such as the Froude number,

$$Fr = \vec{v} \times \vec{v} / \vec{F_G}$$

Rossby number,

$$Ro = \vec{v} \times \vec{v} / \vec{F_C}$$

etc. must be larger than critical values in order for the eddy-like flows of turbulence to take place. This is the basis of our definition of turbulence; that is, **turbulence is defined as an eddy-like state of fluid motion where the inertial-vortex forces of the eddies are larger than any other forces which tend to damp the eddies out**.

The shear instability shown in Fig. 1a applies to perturbations at all scales, but we see that if the velocity amplitude of the perturbations is independent of scale that the overturn time $= L/V$ will be less for small L so that the small eddies will form first and the turbulence cascade will be from small scales to large. The smallest scale eddy is limited by the critical Reynolds number to length scales of about 5 Kolmogorov scales or larger, as shown in Fig. 1a.

A widely quoted misconception about turbulence is that it cascades from large scales to small, but it can easily be seen by observing the growth of turbulent flows such as jets, wakes, and boundary layers, that this is not true. The origin of this belief may be that some turbulent flows extract energy from ambient shears, as shown in Fig. 1a. However, if the external flow is irrotational then its inertial-vortex forces are zero and the flow is nonturbulent, by definition. Its energy cascade from large scales to small is thus nonturbulent. Figure 1b shows an example of a turbulence cascade from small scales to large, where a small rotating cylinder is the source. The cylinder speed is such that the viscous scale $5L_K$ is equal to the cylinder diameter, where $L_K$ is the Kolmogorov scale $(\nu^3/\varepsilon)^{1/4}$.

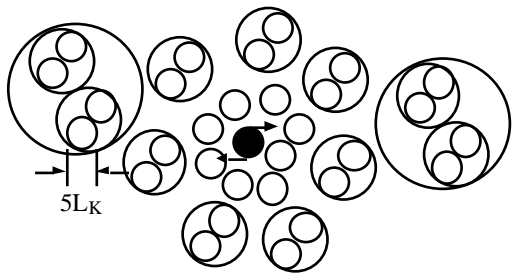

**Fig. 1b. Large scale turbulence produced by a small rotating source**

As shown in Fig. 1b, a series of shear layers form concentric to the rotating cylinder that are unstable and break up to form larger and larger eddies.

## DEFINITION OF FOSSIL TURBULENCE

A very important characteristic of turbulence is that it produces highly persistent, irreversible effects on a wide variety of hydrophysical fields. Linear waves come and go without leaving any trace, but turbulence is intrinsically irreversible and leaves fossil turbulence signatures whenever it happens, especially in natural flows where turbulence is damped out at its largest scales. Familiar examples include the contrails of jet aircraft in the stratified atmosphere. Buoyancy forces damp the large scale turbulence within a few airplane lengths, but the contrails persist and preserve information about the turbulence that produced them. For example, the vertical scale of the contrail is limited by the fossil Ozmidov scale

$$L_{contrail} \quad (L_R)_o = \left[\frac{\varepsilon_o}{N^3}\right]^{1/2}$$

where the stratification frequency $N = (\partial \rho / \partial z \cdot g/\rho)^{1/2}$, z is down, g is gravity, and $\varepsilon_o$ is the viscous dissipation rate of the turbulence at beginning of fossilization, where the Froude number Fr becomes critical. Thus a measurement of $L_{contrail}$ gives an estimate of $\varepsilon_o$. Most patches of temperature, salinity, and density microstructure in the ocean are not actively turbulent at large scales because of fossil turbulence formation, Gibson 2001, 1999. The thickness of spiral galaxy disks may also reflect the fossil Ozmidov scale for turbulence produced by the short-lived supernova stars permitted by the large turbulent Schwarz scales $L_{ST} > L_{SV} > L_{SD}$. In the cold arms of present spiral galaxies, the Jeans scale $L_J$ is generally smaller than the turbulent Schwarz scale in star forming regions, which explains why the star formation rate is much slower than expected using the Jeans criterion.

With strong forcing, turbulence in a rotating, stratified medium grows to a maximum vertical scale determined by $L_R$ and then grows to larger scales in the horizontal until limited by Coriolis forces at $L = (\varepsilon/\Omega^3)^{1/2}$. In the present context, turbulence may be fossilized if the largest scales grow larger than the horizon scale $L_H$, because this is the scale of causal connection. It is necessary to turbulence dynamics that information has time to be transmitted at light speeds to all scales of the turbulence.

Our definition of fossil turbulence follows from the definition of turbulence above. **Fossil turbulence is defined as a fluctuation in any hydrophysical field produced by turbulence that persists after the fluid is no longer turbulent at the scale of the fluctuation.** Such fluctuations include vector fields such as the vorticity $\vec{\omega}$, or the space curvature tensor. Fossil temperature fluctuations produced by turbulence that persist after the turbulence has been damped by inflation stretching space beyond the scale of causal connection are called fossil temperature turbulence, for example.

## BIG BANG TURBULENCE

Turbulence was possible at the time of the big bang because it was so hot. The Planck temperature $T_P = (c^5h/Gk^2)^{1/2} = 1.4 \times 10^{32}$ K exceeds the strong force freeze out temperature $T_{SF} = 10^{28}$ K by 4 orders of magnitude, so there was a brief period between the Planck time $t_P = (c^{-5}hG)^{1/2} = 5.41 \times 10^{-44}$ s and the strong force freeze out time $t_{SF} = 10^{-35}$ s before quarks could form with their gluon force carriers (bosons) to damp out turbulence just as photons damp out turbulence in electron ion plasmas.

General relativity theory breaks down at the Planck length scale $L_P = (c^{-3}hG)^{1/2} = 1.62 \times 10^{-35}$ m, predicting infinite energies at smaller scales. Quantum mechanics also breaks down, predicting infinite probability wave functions. This is the scale where the Compton length $L_C = h/mc$ matches the Schwarzschild length $L_S = Gm/c^2$ at the Planck mass $m_P = (ch/G)^{1/2} = 2.12 \times 10^{-8}$ kg. M-superstring (MS) theory is required to reconcile GR and QM theories, by replacing point particles with multidimensional string-branes that vibrate and wind with tensile force $F_P = c^4/G = 1.1 \times 10^{44}$ kg m s$^{-2}$ in 11 dimensions, Greene 1999. QM fails to include gravity, but we see that MS has it built into the string tension. Spin 2 gravitons arise naturally in string theory. MS theory shows that as black holes approach the Planck mass they can undergo a phase change to photons with zero mass and the same mass-energy.

By combining GR and QM, Hawking showed that black holes have a finite temperature and can radiate energy by pair production at their event horizons. Planck mass particles represent the smallest Schwarzschild black holes, and these can undergo phase changes to zero mass photons with only a small Planck entropy production $s_P = k = 1.38 \times 10^{-23}$ kg m$^2$ s$^{-2}$ K$^{-1}$. The Hawking black hole evaporation temperature $T_H = c^3hG^{-1}m^{-1}k^{-1}$ matches the Planck temperature for a black hole of mass $m = m_P$, and the evaporation time is the Planck time $t_P$.

We can explain the first step of the big bang mechanism as quantum tunneling of a pair of Planck particles that have emerged from the vacuum long enough to raise the temperature for a brief instant and produce a few more. Such events are permitted by the Heisenberg uncertainty principle of QM

$$h \approx E \times t$$

for the maximum existence time $t$ of virtual particles with energy $E$ permitted by vacuum oscillations. If $E$ is the Planck energy $E_P = (c^5hG^{-1})^{1/2}$ then the particle-antiparticle pair can exist no longer than the Planck time $t_P$ without entropy production. With a number of such particles in random interaction it is clear that space, time, and entropy will be produced until the particles and antiparticles find each other and annihilate.

A second critical step of the big bang process occurs if a Planck particle-antiparticle pair become misaligned as they collapse to form an extreme Planck-Kerr black hole, with spin. Planck particles and antiparticles are extreme Schwarzschild black holes, with no spin. This first stable quantum state of the Planck particle fluid is proposed in the present model to be the equivalent of positronium, which is a spinning electron-positron pair formed during the episode of intense electron positron pair production that occurs at the much lower temperature of supernovas, about $10^9$ K. The idea is illustrated in Figure 2. Vacuum oscillations to form Planck particles and Planck antiparticles reversibly is shown on the left, leading to an irreversible escape. More Planck particles form and most are annihilated, but a truly explosive result can occur if a Planck-Kerr particle forms, since one of these can trigger a big bang turbulence cascade by its powerful interactions with Planck particles accreting in the prograde direction. Note that small asymmetries are introduced in the universe vorticity field by the first few Planck Planck-Kerr interactions that may account for the subsequent matter versus antimatter asymmetry.

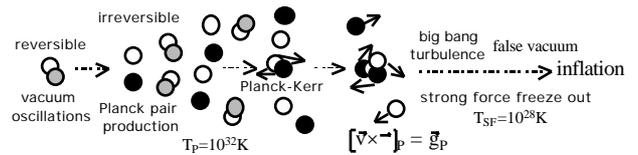

**Fig. 2. Big bang turbulence**

The force attracting Planck particles to Planck antiparticles is gravity. Because they are so close together, the Planck gravitational acceleration $g_P = (c^7h^{-1}G^{-1})^{1/2} = 5.7 \times 10^{51}$ m s$^{-2}$ is enormous. However, when Planck particles are attracted by gravity to spinning Plank-Kerr particles and approach in a prograde direction, Planck inertial-vortex forces appear with the same order of magnitude and opposite direction as the Planck gravity, and these tend to rip the structure apart or prolong its lifetime. The metric describing a nonrotating black hole was discovered by Karl Schwarzschild in 1916, within a year of Einstein's publication of his theory of general relativity. Black holes with rotation are called Kerr black holes because their metric was discovered by R. P. Kerr in 1963, Peacock 2000.

A unique aspect of Kerr black holes is the large amount of energy released as they absorb matter, especially if the matter approaches the spinning black hole in a prograde orbit. Nuclear reactions of stars release only 0.7% of the rest mass energy of hydrogen as it fuses to form helium, but a prograde Kerr absorption of matter can release 42% of the rest mass energy or more, Peacock 2000, p61. The process of radiation by prograde matter accretion is the basis of our big bang turbulence model. The fraction of the rest mass energy $mc^2$ released as radiation when mass $m$ is accreted by a black hole is determined by the size of the marginally stable orbit. The smaller the orbit the greater the fraction of energy radiated. The marginally stable orbit for Schwarzschild black holes is three times the Schwarzschild radius $r_S = 2GM/c^2$, but it is only $r_S/2$ for prograde orbits around Kerr black holes. This accounts for the large accretion efficiency of 42% for prograde accretion for extreme Kerr black holes and only 6% for Schwarzschild black holes and 4% for retrograde accretion by extreme Kerr black holes.

Thus the formation of Planck-Kerr particles greatly increases the probability that a big bang event will occur. Prograde accretion increases the probability of additional Planck pair production by radiation of rest mass energy and focuses the radiation in the direction tangent to the spinning Planck-Kerr. Not only is more energy-space-time produced, but more angular momentum and circulation is produced matching that of the source. Planck-Kerr particles and the more complex Kerr objects with other Planck particles in orbit (see Fig. 2) form powerful pinwheel-like sources of more Planck particles with the same vorticity. These new Planck particle pairs develop random directions and thus cool as the big bang universe expands. All the ingredients of a turbulence cascade are present. Turbulence in the Planck particle gas smoothes fluctuations in the temperature according to the usual rules of turbulent mixing, in this case between the Planck temperature of $10^{32}$ K and the strong force freeze out temperature of $10^{28}$ K. A close up picture of the mechanism is shown in Figure 3. The yin-yang object in the center represents a Planck-Kerr particle spinning in the clockwise direction, spewing out Planck particle gas as two particles attempt to take up prograde orbits. Some of these will take up outer orbits or form Planck-Kerrs. Some will form Planck gas turbulence.

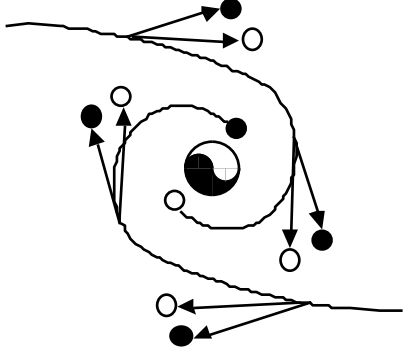

**Fig. 3. Planck pair production by prograde accretion of Planck particles on a spinning Planck-Kerr**

Turbulence can form in the Planck particle gas produced by the Planck-Kerr particles because initially all the gas is so hot that no other particles are possible, other than magnetic monopoles toward the end of the big bang process with mass $10^{-10}$ kg, Guth 1997. Viscous forces are supplied by gravity at the Planck scale, giving a Planck kinematic viscosity $\nu_P = (c^{-1}hG)^{1/2} = 4.9 \cdot 10^{-27}$ m$^2$ s$^{-1}$ and a maximum Reynolds number $Re_{SF} = L_{SF}v_{SF}/\nu_P = 10^6$, where $L_{SF}$ is the strong force horizon $3 \cdot 10^{-27}$ m and the strong force velocity $v_{SF}$ is $(T_{SF}/T_P)^{1/2}$ c = $10^{-2}$ c. This powerful turbulence homogenizes the temperature, vorticity, and space-time-energy as it is produced, with Planck power $P_P = 3.64 \cdot 10^{52}$ kg m$^2$ s$^{-3}$; that is, $10^4$ times larger than the power radiated by all the stars within our present horizon. The cascade is from small scales to large, as shown in Figure 4, driven by the rotating Planck gas expelled by Planck-Kerr particles as shown in Fig. 3. Because no other particles exist besides Planck particles and antiparticles, both momentum and heat are transported by the same mechanism so the Prandtl number is one.

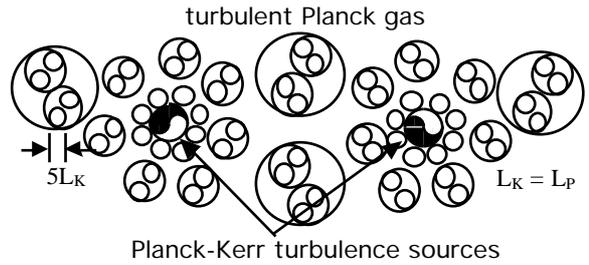

**Fig. 4. Plank-Kerr particles drive a Planck gas turbulence cascade to larger scales**

The big bang starts with the Planck density $\rho_P = c^5h^{-1}G^{-2} = 5.4 \cdot 10^{96}$ kg m$^{-3}$ and decreases as the size of the universe increases due to big bang turbulence production of energy-mass-space-time and more Planck gas. After about $10^{-35}$ s the temperature should decrease to about $10^{28}$ K between the Planck-Kerr particles. Negative pressures are produced by the inertial-vortex forces, the false vacuum, and the bulk viscosity of the gluon-quark plasma, Chen and Spiegel 2001, Paul 2001. At some point, exponential inflation begins, as described by Guth, 1997.

A variety of inflationary models exist, but all depend on a remarkable property of Einstein's equations, Peacock 2000, p25

$$G^{ij} + \Lambda g^{ij} = -\frac{8\pi G}{c^4}T^{ij} ; i,j = 0\text{-}3$$

where $G^{ij}$ is Einstein's tensor, $\Lambda$ is the cosmological constant, $g^{ij}$ is the metric tensor, G is Newton's gravitational constant, and $T^{ij}$ is the momentum energy tensor. The remarkable property is that any process that achieves a sufficiently negative pressure and normal stress p, so that $\rho c^2 + 3p < 0$, will produce gravitational repulsion and an exponentially rapid inflation of space. One such process, independently discovered by Guth and Linde, is the false vacuum, where the big bang universe supercools briefly at the strong force freeze out temperature with the Higgs scalar fields greatly displaced from their large equilibrium vacuum potentials by the very high temperatures. Scalar Higgs fields are required in grand unified quantum field theories to permit spontaneous symmetry breaking, and to provide rest mass for electromagnetic, weak field and strong field particles like electrons, neutrinos and quarks. The simplest grand unified theory has 24 Higgs fields. Other QM fields vanish in the vacuum, but Higgs fields do not. The false vacuum has the peculiar property that its mass-energy density remains constant as space expands, in contrast to the true vacuum where energy density decreases as space expands and photons redshift and cool. According to Einstein's equations, a positive pressure results in gravitational attraction and a negative pressure results in gravitational repulsion, which was the reason Einstein invented  . Both the bulk viscosity of the quark-gluon plasma and the appearance of the false vacuum result in large negative vacuum pressures and consequent gravitational repulsion, exponential expansion of space, and reheating back to nearly the initial temperature of $10^{28}$ K at time t = $10^{-33}$ s when the Higgs fields tunnel back to their true vacuum values, Guth 1997, p171.

Without inflation the maximum mass-energy of the universe was at most the universe volume times the Planck density  $10^{25}$ kg (about the mass of the Earth) assuming an ordinary GR expansion of space by a factor of ten larger than the horizon at $10^{-33}$ s. With inflation, the mass-energy increases to more than $10^{75}$ kg assuming a vacuum density of $10^{80}$ kg m$^{-3}$, $10^{22}$ larger than the present horizon mass for the flat universe produced.

During the inflation period, all lengths were stretched by the same large factor. The inflation factor was $10^{25}$ according to Guth, 1997. The turbulent temperature spectrum should be simply stretched intact to wavenumbers smaller by the inflation factor, so that the smallest scale fluctuations at the stretched Planck scale $10^{-10}$ m will be much larger than the horizon scale at the end of inflation $10^{-24}$ m. Thus the temperature fluctuations up to the inflated strong force freeze out scale $10^{-2}$ m represent the first fossil temperature turbulence because all these fluctuations were caused by turbulence but turbulence is no longer possible at these scales because they are outside the horizon scale of causal connection $L_H$.

## DISCUSSION

The evolution of temperature gradient spectra $k^2$ before and after inflation is illustrated by Figure 5.

The expected turbulent temperature spectrum    is the universal Corrsin-Obukhov-Batchelor form between the Planck scale and the strong force freeze out scale

$$= {}^{-1/3}k^{-5/3}$$

where  is a universal constant about 0.5,  is the temperature variance dissipation rate,  is the viscous dissipation rate, and k is the wavenumber. The Prandtl number is one, so the Planck, Kolmogorov, and Batchelor scales coincide on the right of Fig. 5. The temperature gradient spectra have slope 1/3 and widen to 8 decades as space expands and cools, contrary to standard cosmological assumptions that the spectrum should be flat, as for the Harrison-Zel'dovich "scale independent" spectra, or tilted with negative slope.

## COSMIC MICROWAVE BACKGROUND OBSERVATIONS

An additional $10^{25}$ expansion factor occurs from Einstein's equations between the end of inflation to the time of the plasma to gas transition at $10^{13}$ s, forming the cosmic microwave background pattern of temperature fluctuations on the sky. This pattern was first observed from space by the Cosmic Background Explorer (COBE) satellite, launched in 1989, and by a wide variety of observations since, as summarized by Hu 2000 and Figure 6, adapted from Hu's figure. COBE used a 7 degree beam and covered the entire sky. A very interesting Fig. 6 datum is the low-frequency-cut-off COBE point, not shown in Hu's figure, which corresponds to the $10^{50}$ total inflation factor of the strong force freeze out scale shown in Fig. 5.

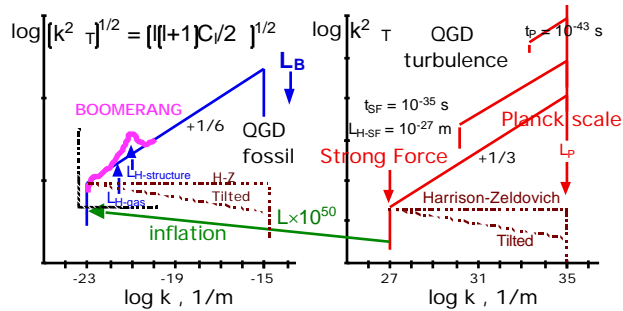

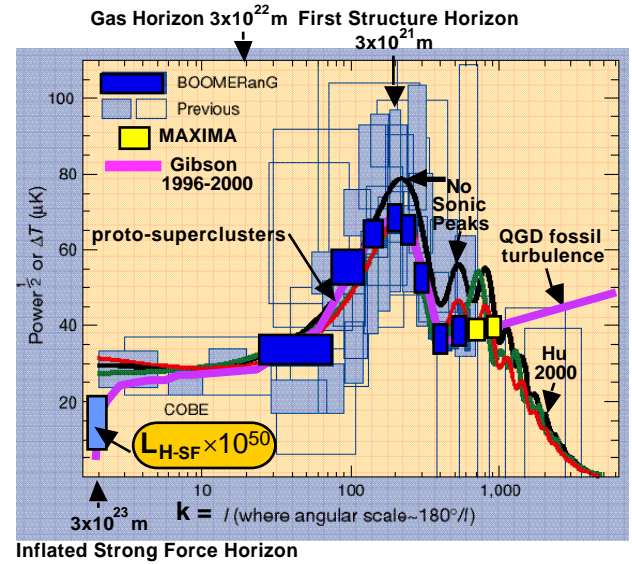

**Fig. 5. Big bang QGD turbulent temperature dissipation spectra on the right are stretched by a factor of $10^{50}$ by inflation and Hubble expansion of space to produce the observed CMB spectrum**

**Fig. 6. CMB spectral observations versus CDM-sonic and Gibson 1996-2000 theories, from Hu 2000 figure.**

The plasma-gas horizon wavelength 3 $10^{22}$ m in Fig. 6 is larger than the wavelength of the peak at 3 $10^{21}$ m, consistent with the interpretation that the peak reflects the wavelength of the horizon at a correspondingly earlier time $10^{12}$ s when decreasing $L_{SV}$ and $L_{ST}$ first matched the increasing horizon $L_H$ so that the first gravitational structures could form, Gibson 1996, 2000. Beyond the primary peak the spectral level returns to a higher level for more than an octave from both the BOOMERANG and MAXIMA observations, consistent with a rising spectal form, such as a $k^{1/6}$ function from the QGD fossil big bang turbulence prediction shown in Fig. 5.

An alternative interpretation of these observations favored by many astrophysical authors is that the primary peak represents (very loud) sounds produced by baryonic matter as it falls into potential wells provided by cold dark matter fragments that serve as nuclei for galaxy formation. Secondary and tertiary sound peaks were not observed by the 2000 reports of the BOOMERANG collaboration, but after careful reanalysis such peaks are claimed in 2001 reports, as shown in Figure 7. An alternative interpretation is shown by the line of Fig. 7, which is that the observed bumps are reflections of gravitational fragmentation of the plasma to form proto-galaxy-clusters and proto-galaxies, not sound. The proto-galaxy wavenumber should be a factor of 10 larger than the primary peak, as shown, to reflect about $10^{-3}$ smaller mass.

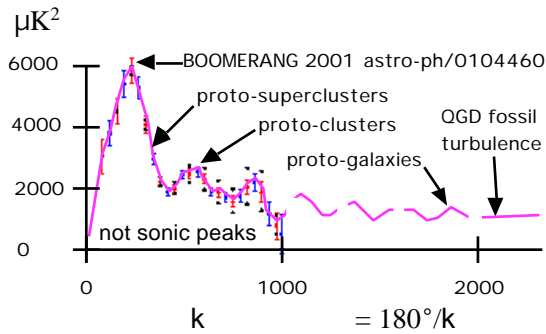

**Fig. 7. CMB peaks, Netterfield et al. 2001.**

Sonic models are all based on the Jeans 1902 gravitational condensation criterion, which is invalid according to Gibson 1996, 2000. Nonbaryonic dark matter is too diffusive to form fragments before $10^{13}$ s, and cannot form fragments to support sonic oscillations.

Future measurements of the CMB are planned with the Microwave Anisotropy Probe (MAP) satellite, which is now going into orbit about the L2 Lagrange point located $1.5 \ 10^9$ m beyond the earth away from the sun. MAP measurements should greatly improve the full sky coverage of the CMB and the precision of its temperature spectral shape. Assuming the high wavenumber portion of the spectrum is due to big bang fossil turbulence, averaging over the full sky is needed to compensate for high levels of intermittency expected for large Reynolds number turbulence and turbulent mixing, from Kolmogorov's third universal similarity hypothesis. Intermittency can lead to underestimates due to undersampling errors. Figure 8 shows a simulated map of temperature fluctuations on the sky from MAP after correcting for the dipole error due to motions toward the great attractor, but not removing the central band of microwave noise sources on the plane of the Galaxy.

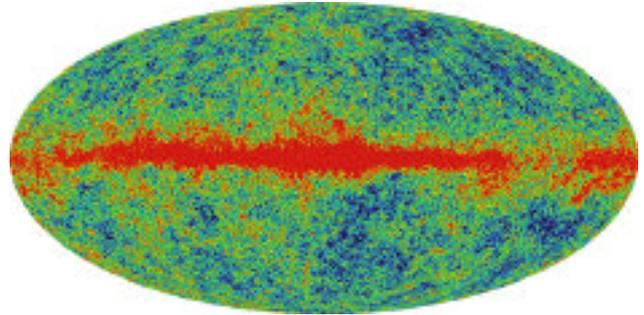

**Fig. 8. Simulated MAP CMB temperature fluctuations, http://map.gsfc.nasa.gov.**

The all sky point from COBE shown on the left of Figure 6 is at spherical wavenumber l = 2, corresponding to an angle of 90°, and is the smallest possible spectral point that can be estimated from the CMB observations. As one can see from Fig. 6, the statistical significance is uncomfortably low for the single COBE point showing a low wavenumber cutoff at the inflated strong force freeze out point, so the improved statistics at large scales expected from MAP observations will be valuable.

## SUMMARY

We have shown that all physical theories are necessary components to a description of the first turbulence and the origin of the universe; that is, general relativity (GR), quantum mechanics (QM), M-superstring theory (MS), and fluid mechanics (FM). As shown in Figure 9, limited scale ranges exist where only one theory is needed to provide satisfactory descriptions, some areas require two or three, but the full description of big bang turbulence and big bang fossil turbulence is an area where none can be omitted. Quantum mechanics describes quantum tunneling to make Planck particle pairs, general relativity describes the Planck versus Planck-Kerr big bang turbulence (inertial-vortex force) interaction, string theory describes the equivalence of Planck particle black holes and photons, and fluid mechanics describes turbulence, turbulent mixing and fossil turbulence phenomena.

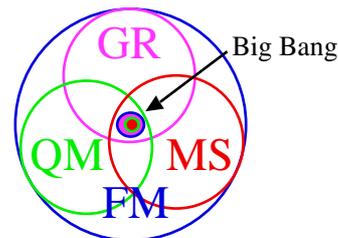

**Fig. 9. Theories required to describe big bang turbulence and big bang fossil turbulence.**

Fluid mechanics is also necessary to exclude misconceptions such as the Jeans 1902 criterion for gravitational structure formation and cold dark matter theories, and to provide revised criteria that take into account turbulence forces, viscous forces, and diffusion velocities, Gibson 1996, 2000. For additional information, see http://www-acs.ucsd.edu/~ir118 with figures and links to related reports in the Los Alamos astro-ph archive.

## CONCLUSIONS

Turbulence is defined in terms of the inertial-vortex force formed by the cross product of velocity and vorticity. By this definition and a new proposed mechanism for the quantum-gravitational-dynamics epoch, turbulence first occurred at the time of the big bang and was responsible for the big bang. Vacuum oscillations and quantum tunneling provide the first Planck particles and antiparticles, but some of these must form extreme, rotating, Planck-Kerr black holes in order to trigger the big bang explosion of space-time-energy production. Prograde accretion of Planck particles on Planck-Kerr black holes produces mass-energy radiation with high efficiency, further Planck pair production, and expansion of the zone of Planck gas turbulence with the same angular momentum as the Planck-Kerr. The inertial-vortex force $(c^7 h^{-1} G^{-1})^{1/2} = 5.7 \times 10^{51}$ m s$^{-2}$ at Planck scales matches the gravitational Planck force, and therefore ranks as a new fundamental fifth force of nature along with the strong, weak, gravitational and electromagnetic forces. It supplies a powerful negative pressure to drive the expansion of space during the initial big bang, and produces the Planck gas turbulence that homogenizes Planck temperatures with the lower temperatures formed by the expansion. Fossils of big bang turbulent temperature mixing are preserved by a variety of features of the cosmic microwave background spectrum.

## ACKNOWLEDGEMENTS

The author would like to thank Professor Fazle Hussain for discussions and comments leading to this paper.

## REFERENCES

Binney, J. and S. Tremaine, "Galactic Dynamics", Princeton University Press, New Jersey, 1987.
Chen, X. and E. A. Spiegel, Radiative bulk viscosity, Mon. Not. R. Astron. Soc., 324, 865-871, 2001.
Gibson, C. H., Kolmogorov similarity hypotheses for scalar fields: sampling intermittent turbulent mixing in the ocean and Galaxy, in "Turbulence and Stochastic Processes: Kolmogorov's Ideas 50 Years On", Proceedings of the Royal Society London, Ser. A, V434 N1890, 149-164, 1991.
Gibson, C. H., Turbulence in the ocean, atmosphere, galaxy, and universe, Applied Mechanics Reviews, 49:5, 299-315, 1996.
Gibson, C. H., Fossil turbulence revisited, Journal of Marine Systems, vol. 21, nos. 1-4, 147-167, 1999.
Gibson, C. H., Turbulent mixing, diffusion and gravity in the formation of cosmological structures: the fluid mechanics of dark matter, Journal of Fluids Engineering, 122, 830-835, 2000.
Gibson, C. H., Fossil Turbulence, "Encyclopedia of Ocean Sciences", Academic Press, Ed. J. Steele, 2001.
Greene, Brian, "The Elegant Universe", Norton, NY, 1999.
Guth, Alan, "The Inflationary Universe", Addison Wesley, 1997.
Hu, Wayne, Ringing in the new cosmology, Nature, 404, 939, 2000.
Jeans, J. H., The stability of a spherical nebula, Phil. Trans. R. Soc. Lond. A, 199, 1, 1902.
Kolb, E. W. and M. S. Turner, "The Early Universe", Addison Wesley, NY, 1990.
Larsen, S. S., J. P. Brodie, B. G. Elmegreen, Y. N. Efremov, P. W. Hodge, and T. Richtler, Structure and mass of a young globular cluster in NGC 6946, Astrophysical Journal, 556, 801-812, 2001.
Netterfield, C. B., P. A. R. Ade, J. J. Bock, and 24 others, A measurement by BOOMERANG of multiple peaks in the angular power spectrum of the cosmic microwave background, astro-ph/0104460v2, 2001.
Ozernoy, L. M. and G. V. Chibisov, Galactic Parameters as a Consequence of Cosmological Turbulence, Astrophys. Lett. 7, 201-204, 1971.
Paul, B. C., Viscous cosmologies with extra dimensions, Physical Review D, 64, 027302-3, 2001.
Peacock, J. A., "Cosmological Physics", Cambridge University Press, 2000.
Padmanabhan, T., "Structure Formation in the Universe", Cambridge University Press, Cambridge UK, 1993.
Peebles, P. J. E., "Principles of Physical Cosmology", Princeton University Press, Princeton, NJ, 1993.
Rees, Martin, "New Perspectives in Astrophysical Cosmology", Cambridge University Press, UK, 2000.
Schild, R. E., Microlensing variability of the gravitationally lensed quasar Q0957+561 A,B, Astrophysical Journal, 464, 125-130, 1996.
Silk, Joseph, "The Big Bang", W. H. Freeman and Company, NY, 1989.
Weinberg, S., "Gravitation and Cosmology: Principles and Applications of the General Theory of Relativity", John Wiley & Sons, New York, 1972.
Weinberg, S., "The First Three Minutes", Basic Books, Inc., Publishers, New York, 1977.